\documentclass[12pt,preprint]{aastex}

\usepackage{bm}
\usepackage{epstopdf} 
\usepackage{natbib}
\bibpunct{(}{)}{;}{a}{,}{,}
\usepackage[backref,colorlinks,citecolor=blue]{hyperref}
\usepackage{threeparttable}
\usepackage{booktabs}

\shorttitle{NIR spectra of DY\,Persei stars} 
\shortauthors{Anirban, Pandey, Joshi \& Ashok}

\begin{document}
\title{Are DY\,Persei stars cooler cousins of R Coronae Borealis stars?}

\author{Anirban Bhowmick$^{1}$, Gajendra Pandey$^{1}$, Vishal Joshi$^{2}$ and N.M.
Ashok$^{2}$} 
\affil{$^{1}$Indian Institute of Astrophysics, Koramangala, Bengaluru
 560 034, India;\\
$^{2}$Physical Research Laboratory, Ahmedabad 380009, India;\\}

\email{anirban@iiap.res.in; pandey@iiap.res.in; vjoshi@prl.res.in; 
ashok@prl.res.in}

\begin{abstract}
In this paper we present for the first time, the study of low resolution  $H$- and $K$- band spectra of 7 DY\,Per type and suspects stars as well as DY\,Persei itself. We also observed $H$- and $K$- band spectra of 3 R Coronae Borealis (RCB) stars, 1 hydrogen-deficient carbon (HdC) star and 14 cool carbon stars including normal giants as comparisons. High $^{12}$C/$^{13}$C and low $^{16}$O/$^{18}$O ratios are characteristic features of majority RCBs and HdCs. We have estimated $^{16}$O/$^{18}$O ratios of the programme stars from  the relative strengths of the $^{12}$C$^{16}$O and $^{12}$C$^{18}$O molecular bands observed in $K$- band. Our preliminary analysis suggest that a quartet of the DY\,Per suspects along with DY\,Persei itself seems to show isotopic ratio strength consistent with the ones of RCB/HdC stars whereas two of them do not show significant $^{13}$C and $^{18}$O in their atmospheres. Our analysis provides further indications that DY\,Per type stars could be related to RCB/HdC class of stars.
\end{abstract}
\section{INTRODUCTION}
R Coronae Borealis (RCB) stars are low mass, hydrogen deficient carbon rich yellow supergiants associated with very late stages of stellar evolution.
These are characterised by their unusual light variability showing a rapid aperiodic light dimming of several magnitudes in the optical with a slow return to their maximum light, and exhibit IR excess \citep{hz-PG1938,hz-C1996}. 

Six hydrogen deficient carbon stars (HdCs) are known. They are spectroscopically similar to RCBs but most of them do not exhibit light declines or show IR excess \citep{hz-W1967,hz-C2012}, the exception being HD\,175893 that shows IR excess \citep{hz-T2012}. 

DY\,Persei and DY\,Persei type (DY\,Per type) stars are however, a peculiar class of cooler carbon stars showing also dramatic but slower light declines than RCBs and with more symmetric rise in time. Some IR excess \citep{hz-Ak1994,hz-A2001} is also observed for these stars with a somewhat warmer circumstellar shells than RCBs \citep{hz-T2009}. DY\,Per type star candidates are the stars having similar light curve and position in $J-H$ and $H-K$ diagram like DY\,Per type stars found so far, but without any spectroscopic observations or confirmations.
We introduce here the term DY\,Per suspect, i.e. carbon stars showing spectroscopic features similar to DY\,Per type stars but whose light curve has not shown characteristic symmetrical decline events but rather large photometric variations that could also be due to dust obscuration.

The effective temperatures of DY\,Per type stars appear to be at the
cooler end of the known RCB stars \citep{hz-KB1997}. DY\,Per type stars may be hydrogen deficient due to absence of hydrogen Balmer lines in their spectra, 
nevertheless, the status of hydrogen deficiency is not yet clear due to their cooler effective temperatures and absence of flux in the $G$ band of CH at 4300\AA\ region \citep{hz-KB1997,hz-Z2007,hz-Y2009}. Until now in addition to DY\,Persei itself, only seven Galactic DY\,Per type stars are known \citep{hz-T2008,hz-T2013,hz-M2012}.
\citet{hz-A2001} and \cite{hz-T2004,hz-T2009} reported around 27 Magellanic DY\,Per type stars and candidates with more possible suspects given by \citet{hz-S2009} through their OGLE-III light curves.
Due to the small number of known DY\,Per type stars and candidates, it is a challenge to characterise these stars and investigate any possible connection with the RCBs. We therefore also introduced DY\,Per suspect stars in our study.

Two scenarios have been proposed to explain the evolutionary origin of an RCB star: first, the double-degenerate merger (DD) scenario involving 
the merger of an He and a C-O white dwarf, \citep{hz-W1984,hz-SJ2002,hz-Pet2006}, and second, the final helium shell flash (FF) \citep{hz-I1983}, scenario involving a single star evolving into planetary nebular (PN) phase or post asymptotic giant branch (post-AGB) phase contracting towards the white dwarf sequence. The ignition of the helium shell in a post-AGB star, say, a cooling white dwarf, results in what is known as a late or very late thermal pulse \citep{hz-H2001} that injests the thin hydrogen rich outer layer, making the star hydrogen deficient, and the star expands to supergiant dimensions \citep{hz-F1977,hz-R2007}.

Based on the fluorine \citep{hz-P2008} , $^{13}$C \citep{hz-H2012}, and $^{18}$O \citep{hz-C2005,hz-C2007,hz-G2009,hz-G2010} abundances in RCB and HdC stars, a consensus is now emerging for DD scenario, however,
a small fraction of these may be produced by FF scenario \citep{hz-C2011}.


\begin{table*}[ht]
\caption{Log of observations of RCB and HdC stars as well as DY\,Persei and the DY\,Per affiliated stars.}
\centering
\small
\begin{threeparttable}
\begin{tabular}{lllll}\hline\label{Table-1}
Star name  & Date of observation & $K$- mag.\tnote{*} & S/N & Star type\\ 
(SIMBAD) & & (SIMBAD) & (2.29 \textit{$\mu$}) & \\ \hline

HD 137613  & 16 April 2016 & 5.25 & 70 & HdC \\

Z\,UMi &  01 May 2016 & 7.3 & 55 & RCrB \\ 

SV\,Sge &  18 November 2016 & 5.9 & 110 & RCrB \\

ES Aql  & 18 November 2016 & 7.9 & 105 & RCrB \\ 

DY Persei  & 04 October 2014, 16 January 2016, & 4.4 & 105 & DY\,Per prototype\\ 
 & 17 Jan 2016 , 06 November 2016 & \\

ASAS J065113+0222.1  & 16 January 2016, 17 Jan 2016 , & 4.9 & 80 & DY\,Per type star\tnote{a}\\
& 23 February 2017 & \\

ASAS J040907-0914.2 & 16 January 2016, 17 Jan 2016 , & 3.6 & 95  & DY\,Per suspect\tnote{b}\\ 
(EV Eri) & 06 November 2016, 23 February 2017 & \\

ASAS J052114+0721.3  & 16 January 2016, 06 November 2016, & 2.19 & 110 & DY\,Per suspect\tnote{b}\\ 
(V1368 Ori) & 23 February 2017 & \\   

ASAS J045331+2246.5 & 04 October 2014, 16 January 2016, & 2.84 & 80 & DY\,Per suspect\tnote{b} \\ 
 & 17 Jan 2016 &  \\

ASAS J054635+2538.1  & 16 January 2016, 17 Jan 2016 , & 4.3 & 90 & DY\,Per suspect\tnote{b} \\
 (CGCS 1049)& 18 March 2016 & \\ 
 
ASAS J053302+1808.0 & 16 January 2016, 17 Jan 2016 & 5.6 & 90 & DY\,Per suspect\tnote{b}\\ 
(IRAS 05301+1805) &  &  &   & \\

ASAS J191909-1554.4 & 01 July 2016 & 1.06 & 105   & DY\,Per type star\tnote{a} \\ 
(V1942\,Sgr ) &  &  &   & \\ \hline

\end{tabular}
\begin{tablenotes}
\item [a] \citet{hz-M2012}
\item [b] \citet{hz-T2013}
\item [*] \textit{Reported from the Two Micron All Sky Survey Point Source Catalogue \citep{hz-Cu2003}}
\end{tablenotes}
\end{threeparttable}
\end{table*}


Alongwith hydrogen-deficiency, the main spectral characteristics of RCBs and HdCs that distinguishes them from normal AGB and post-AGB stars are the presence of very high amounts of $^{18}$O and weak or no presence of $^{13}$C in their atmospheres. Using the NIR, $K$-band spectra of these stars, \citet{hz-C2005,hz-C2007,hz-G2009,hz-G2010} found that the isotopic ratios of $^{16}$O/$^{18}$O, derived from the relative strengths of the observed $^{12}$C$^{16}$O and $^{12}$C$^{18}$O molecular bands, range from 0.3 to 20. Note that the typical value of $^{16}$O/$^{18}$O $\sim$ 500 in solar neighbourhood and 200 to 600 in Galactic interstellar medium \citep{hz-G2002}. Also, the $^{12}$C/$^{13}$C ratio for several RCBs and all HdCs are significantly higher than the CN-equilibrium value of 3.4 \citep{hz-A2001,hz-H2012}. Thus, the low values of $^{16}$O/$^{18}$O and high values 
of $^{12}$C/$^{13}$C in both HdCs and RCBs make it obvious that these two classes of carbon-rich and hydrogen poor stars are indeed closely related.

On the contrary, the possible evolutionary connection of DY\,Per type stars with RCBs/HdCs or with normal carbon rich AGBs needs to be explored.
\citet{hz-Z2007} reported the high resolution spectrum of the DY\,Persei showing significant hydrogen deficiency with high $^{12}$C/$^{13}$C ratio like most RCBs. It is to be noted that the low resolution spectra of DY\,Per type variables in the Magellanic clouds show significant enhancement of $^{13}$C from the isotopic Swan bands at about 4700\AA, but the $^{13}$CN band at 6250\AA\ is not seen \citep{hz-A2001,hz-T2009}. Also, the enhancement of $^{13}$C in the atmospheres of Magellanic DY\,Per type stars is reported for only 9 cases out of 27 \citep{hz-A2001,hz-T2004,hz-T2009}. Hence, there seems to exist a mixed $^{12}$C/$^{13}$C isotopic ratio in Magellanic DY\,Per type stars.

In this paper we search for the contributing spectral features involving $^{18}$O and $^{13}$C in the low resolution $H$- and $K$-band NIR spectra of the observed DY\,Per type stars and DY\,Per suspects. Note that our DY\,Per suspects are the cool carbon stars taken from Table-5 of \citet{hz-T2013} which they rejected as RCB candidates due to enhanced $^{13}$C in their spectra and no clear rapid decline events in their light curves. However, we selected these stars based on their similarity with DY\,Per type stars as  given in the description by \citet{hz-T2013} in their text, verbatim, ``Their light curves show variations up to 2 mag, but with no clear signs of a fast decline. Because they all present large photometric oscillations of $\sim$ 0.8 mag amplitude and their spectra do not show clear signs of presence of hydrogen, they should be considered as DY\,Per star candidates."

The objective is to explore possible connections between DY\,Per type stars and DY\,Per suspects with classical carbon stars or with RCBs/HdCs. Our observations, analysis, and results are discussed in the following Sections.

\section{OBSERVATIONS AND REDUCTIONS}
$H$- and $K$-band spectra of our target stars were obtained from TIFR Near Infrared Spectrometer and Imager (TIRSPEC) \citep{hz-N2014} mounted on 
Himalayan Chandra Telescope (HCT) at Hanle, Ladakh, India. The log of observations is given in Table \ref{Table-1} for the RCB and HdC stars as well as all the DY\,Per affiliated stars, and in Table \ref{Table-2} for the normal and cool carbon stars.


\begin{table*}[ht]
\centering
\caption{Log of observations of normal cool giants selected from \citet{hz-J1992,hz-TA2007}}

\begin{tabular}{llllr}\hline\label{Table-2}
Star name & Date of observation & $K$- mag. & S/N & Star type \\ 
(SIMBAD) &  & (SIMBAD) & (2.29 \textit{$\mu$})& \\ \hline

Arcturus & 01 May 2016 & -2.9 & 85 & K \\ 

HD 156074  & 14 October 2014 & 5.28 & 125 & R  \\ 

HD 112127  & 17 Jan 2016 , 18 March 2016 & 4.17 & 170 & R  \\ 

BD+06 2063  & 16 April 2016 & 4.1 & 205 & S  \\ 

HR 337 & 17 Jan 2016 & -1.85 & 120 & M  \\ 

HD 64332  & 16 April 2016 & 2.3 & 185 & S  \\ 

HD 123821  & 18 March 2016 & 6.3 & 110 & R \\ 

HR 3639  & 16 April 2016 & -1.7 & 130 & S  \\

HD 58521  & 18 March 2016 & -0.44 & 140 & S  \\ 

HD 76846  & 17 Jan 2016 , 18 March 2016 & 6.6 & 130 & R   \\ 

V455\,Pup  & 17 Jan 2016 , 16 April 2016,  & 5.27 & 80 & C  \\
 &  23 February 2017  & \\

TU Gem  & 18 March 2016 & 0.78 & 85 & N \\ 

Y CVn  & 16 April 2016 & -0.81 & 80 & J  \\ 

RY Dra & 16 April 2016 & 0.19 & 75 & J  \\ \hline

\end{tabular}
\end{table*}

Spectra were recorded in cross-dispersal mode in two dithered positions with multiple exposures in each position having average exposure time of 100s for each frames. The frames were combined to improve the signal-to-noise ratio (SNR) (see Tables \ref{Table-1} and \ref{Table-2}). The recorded spectra in the $H$- band appear noisier than the $K$- band  due to lower photon counts.
For stars fainter than $K$-magnitude 6, frames of 500s exposure were taken and combined to improve the SNR. 
After each set of star exposures, three continuum lamp spectra and an argon lamp spectrum were obtained. For removing the telluric lines from the star's spectrum, rapidly rotating O/B type dwarfs (telluric standards) were observed during each observing run in the direction of the programme stars.

The slit setting mode S3 with a slit width of 1.97" was available. For this slit setting the average resolving power at the $H$- and $K$- central wavelength is about $\sim$900 as measured from the FWHM of the clean emission lines of the comparison lamp spectrum.

The data obtained is made available after dark and cosmic ray corrections.  
The Image Reduction and Analysis Facility (IRAF) software package was used to reduce these recorded spectra. The dithered frames of the recorded spectra were combined to correct for background emission lines using the ABBA dithering technique. A master flat was made by combining the continuum lamp spectra. The object frames were flat corrected using standard IRAF tasks. One dimensional (1D) spectrum was then extracted and wavelength calibrated using the argon lamp spectrum. 
The wavelength calibrated star's spectrum is then divided by a telluric standard's spectrum, to remove the telluric absorption lines, using the task TELLURIC in IRAF.

All the 7 DY\,Per affiliated stars (2 DY\,Per type stars and 5 DY\,Per suspects), we observed, were taken from the catalogue of stars presented by \cite{hz-T2013} and \cite{hz-M2012}. Our selection was limited by the location of the Observatory, HCT, where we could observe only the stars north of $-25^{\circ}$ declination.
Three cool RCBs: Z\,UMi, SV\,Sge and ES\,Aql and one HdC star HD\,137613 were also observed. Except for Z\,UMi, the other two RCBs were observed at about their maximum light as verified from the AAVSO database(\href{https://www.aavso.org/}{www.aavso.org}); Z\,UMi was in recovery phase ($\Delta $V$ \sim $3) and so the observed spectrum is particularly noisy. We have also observed a variety of normal giants/supergiants covering the effective temperature range of the programme stars. The normal giants/supergiants were taken from \citet{hz-J1992,hz-TA2007} spanning K giants through N- and J-type cool carbon stars. These stars along with the HdC/RCBs were observed to compare and confirm the presence/absence of $^{13}$C$^{16}$O and $^{12}$C$^{18}$O features in DY\,Per type stars, and  DY\,Per suspects. 

\section{CO BANDS AND OVERVIEW OF THE SPECTRA}
\begin{figure*}
\includegraphics[width=16cm,height=16.cm]{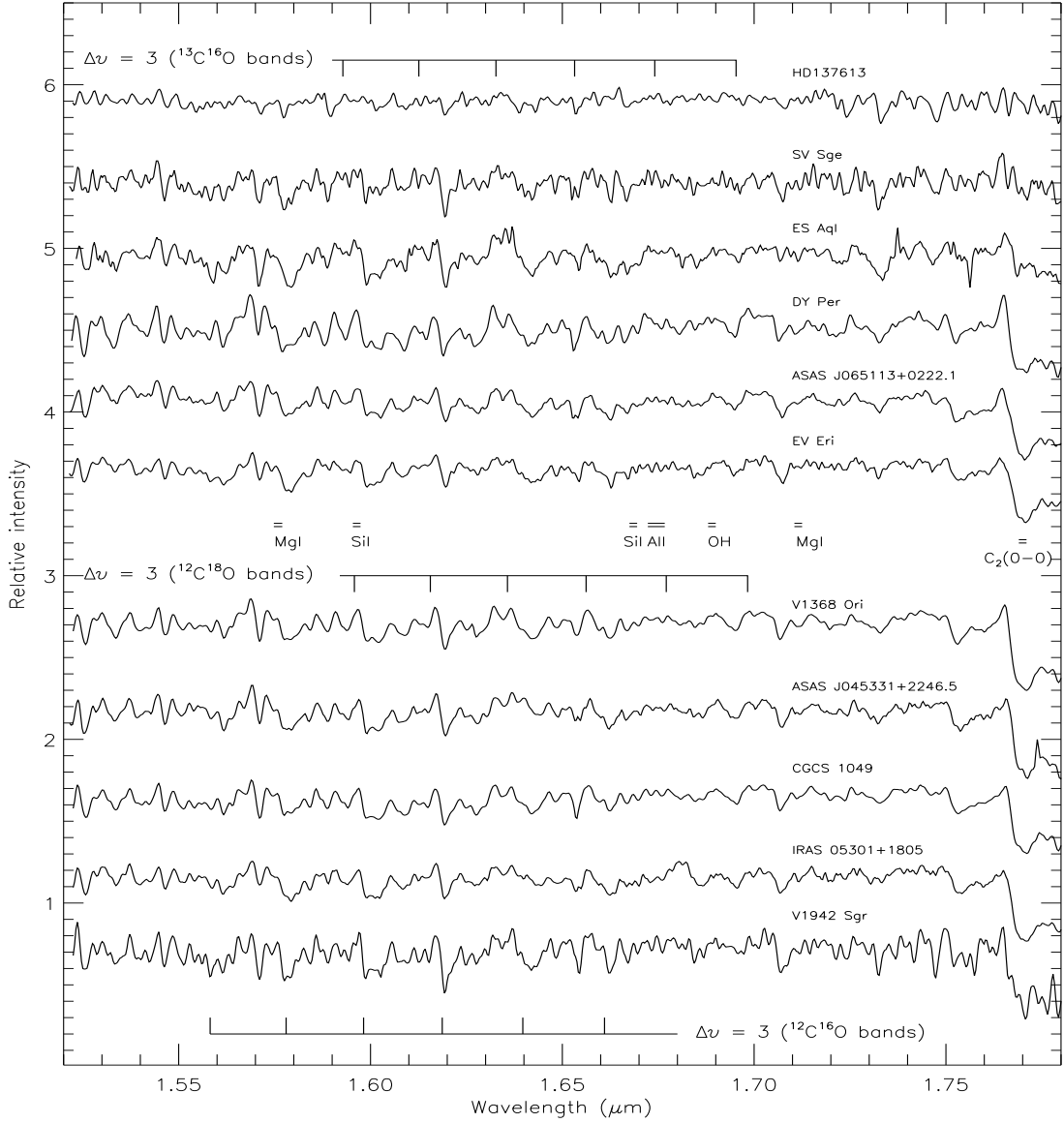}
\caption{1.52$-$1.78 $\mu$m spectra of RCBs, HdCs, DY\,Persei, and DY\,Per affiliated stars. The band head positions of $^{12}$C$^{16}$O ,$^{12}$C$^{18}$O and $^{13}$C$^{16}$O and other key features are marked. The stars are ordered according to their increasing effective temperature (approximate) from bottom to top. }\label{fig.1}

\end{figure*}

\begin{figure*}
\includegraphics[width=16cm,height=16.cm]{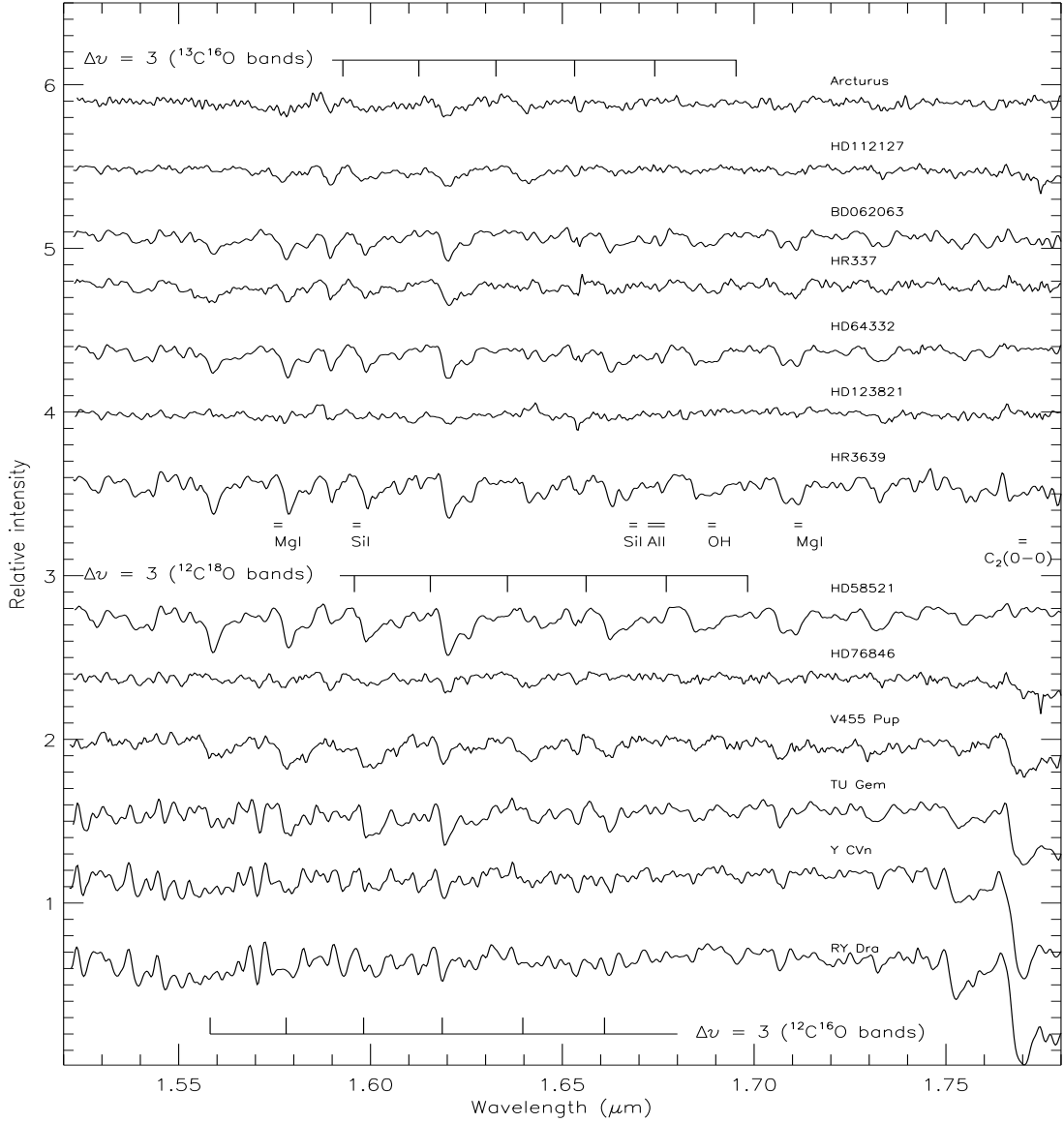}
\caption{1.52$-$1.78 $\mu$m spectra of normal giants/supergiants of different spectral type ranging from K giants on the top to cool N type carbon stars at the bottom. The band head positions of $^{12}$C$^{16}$O , $^{12}$C$^{18}$O and $^{13}$C$^{16}$O and other key features are marked.}\label{fig.2}

\end{figure*}

\begin{figure*}
\includegraphics[width=16.cm,height=16.cm]{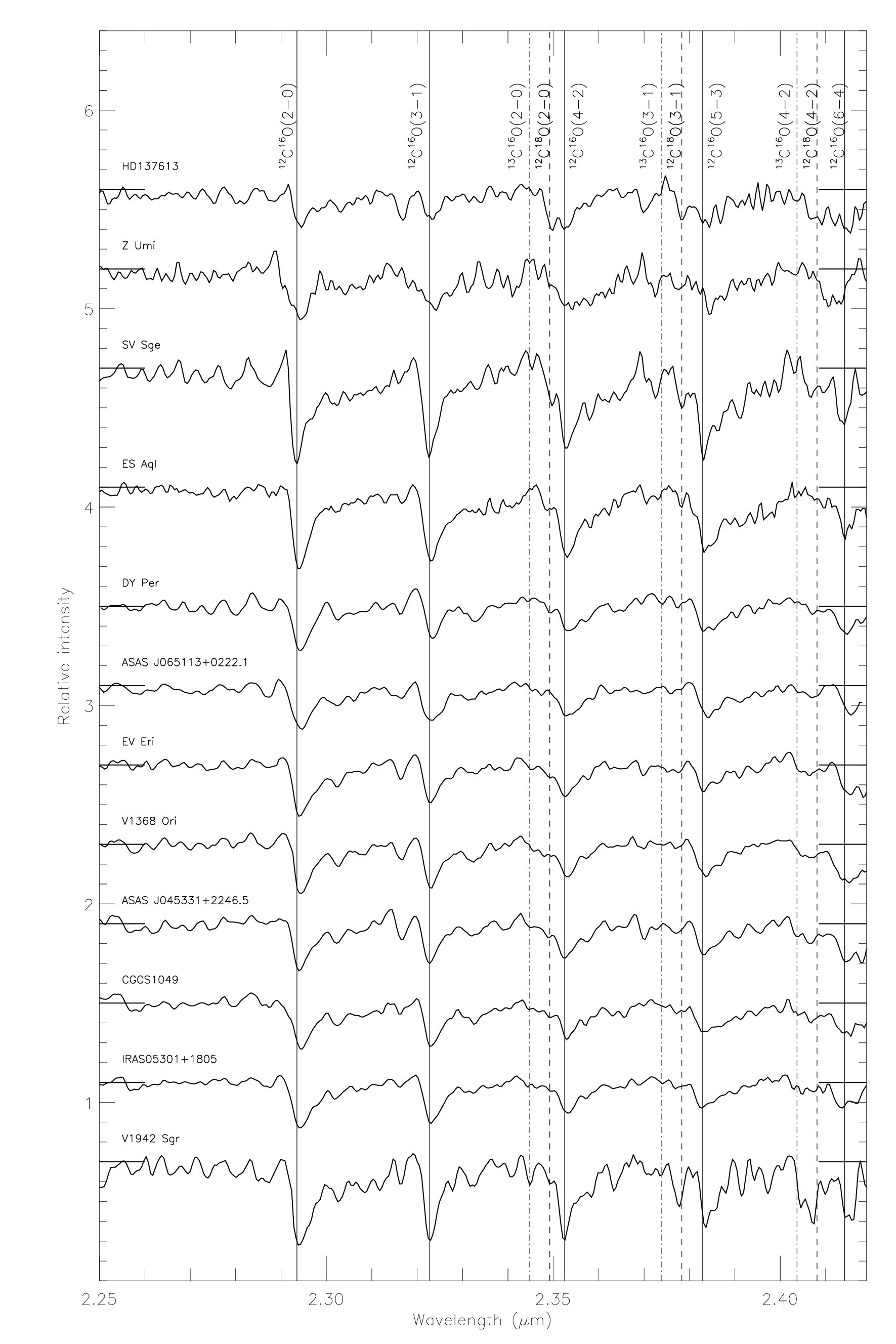}
\caption{2.25$-$2.42 $\mu$m spectra of RCBs, HdC, DY\,Persei,  and DY\,Per affiliated stars , with wavelengths of $^{12}$C$^{16}$O , $^{12}$C$^{18}$O and $^{13}$C$^{16}$O indicated by vertical lines. The stars are ordered according to their increasing effective temperatures (approximate) from bottom to top. The position of the mean continuum for each spectrum is indicated by the line marked. }\label{fig.3}

\end{figure*}

\begin{figure*}
\includegraphics[width=16.cm,height=16.cm]{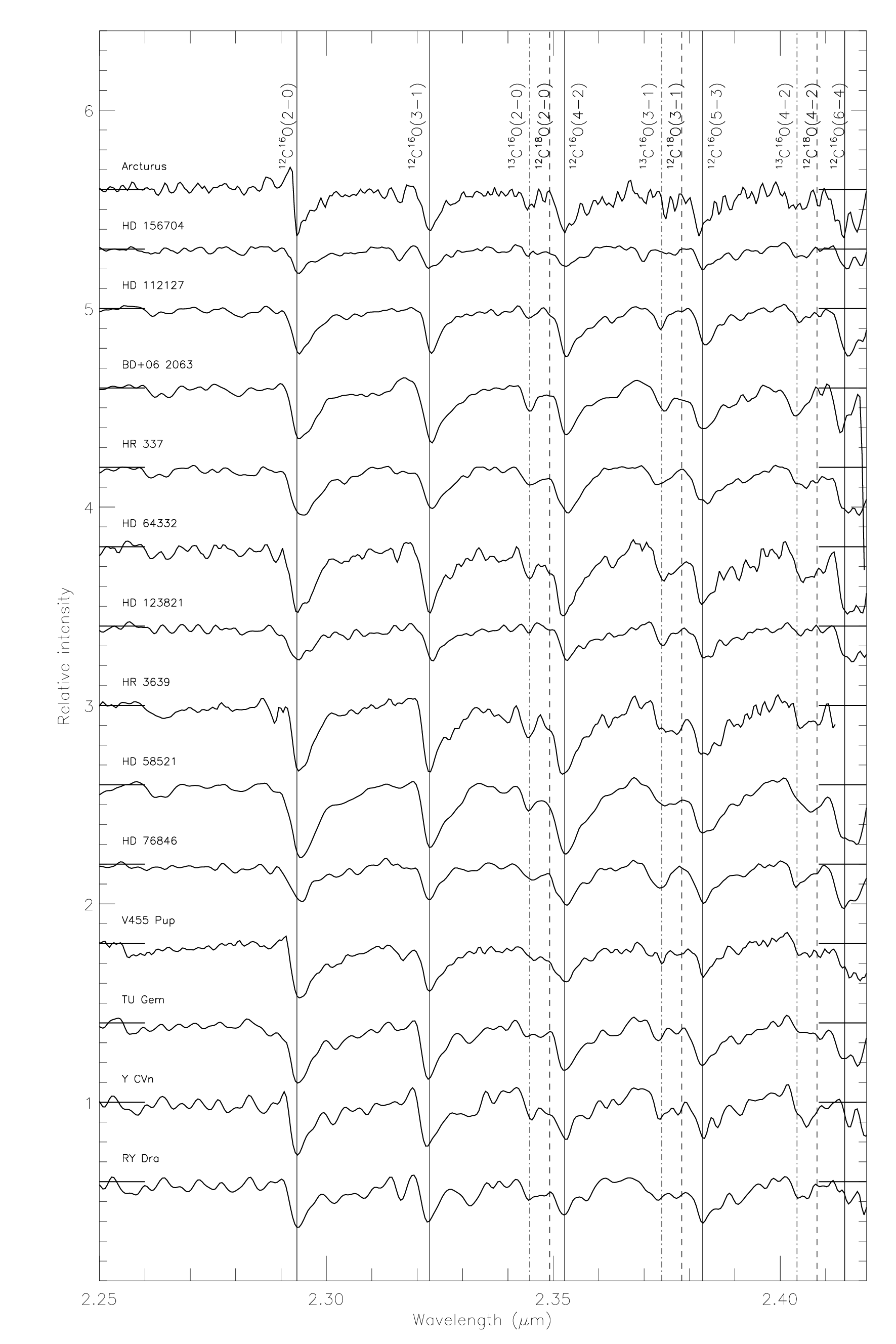}
\caption{2.25$-$2.42 $\mu$m spectra of normal giants of different spectral type ranging from K giants in the top to cool N type carbon stars in the bottom. As in Figure 3 ,wavelengths of $^{12}$C$^{16}$O , $^{12}$C$^{18}$O and  $^{13}$C$^{16}$O indicated by vertical lines. The position of the mean continuum for each spectrum is indicated by the line marked.}\label{fig.4}

\end{figure*}

 The band head wavelengths of $^{12}$C$^{16}$O are available in literature for both $H$- and $K$-band region. We have calculated the wavelengths of  $^{13}$C$^{16}$O and $^{12}$C$^{18}$O by using the standard formula for isotopic shift from \citet{hz-H1950} and the ground state constants of $^{12}$C$^{16}$O are taken from \citet{hz-M1975}. We have verified our calculated band head wavelengths of $^{13}$C$^{16}$O and $^{12}$C$^{18}$O for the first overtone transition, with those given by \citet{hz-C2005} and hence, applied the same procedure to calculate the second overtone band head wavelengths of $^{13}$C$^{16}$O and $^{12}$C$^{18}$O.\\ 
Figures \ref{fig.1} and \ref{fig.2} show the $H$-band (1.52$-$1.78 $\mu$m region) spectra of our programme stars and the comparison stars (normal giants/supergiants), respectively; the 2nd overtone features of $^{12}$C$^{16}$O, $^{12}$C$^{18}$O and $^{13}$C$^{16}$O including C$_{2}$ Balik-Ramsay system (0-0) are marked with other key features. $H$-band spectra of HD\,156704 (normal K giant) and Z\,UMi (RCB) were very noisy, hence, not shown.
The $K$-band (2.25$-$2.42 $\mu$m region) spectra of the programme stars and the comparison stars are shown in Figures \ref{fig.3} and \ref{fig.4}, respectively; the first overtone band heads of $^{12}$C$^{16}$O, $^{12}$C$^{18}$O and $^{13}$C$^{16}$O are indicated. The spectra shown in Figures \ref{fig.1},\ref{fig.2},\ref{fig.3},\ref{fig.4} are normalised to the continuum and are aligned to lab wavelengths of $^{12}$C$^{16}$O band heads. \\


%

\begin{table*}[ht]
\caption{Absorption depths of first overtone CO band heads and the estimated $^{16}$O/$^{18}$O and $^{12}$C/$^{13}$C ratios of RCBs, HdC and DY\,Per affiliate stars. }
\begin{center}
\small

\begin{tabular}{lllllllll}\hline\label{Table.3}

Star name & Star type & \multicolumn{2}{c}{$^{12}$C$^{16}$O} & & \multicolumn{2}{c}{$^{12}$C$^{18}$O}&  $^{16}$O/$^{18}$O  & $^{12}$C/$^{13}$C \\  
\cmidrule{3-4} \cmidrule{6-7} 
 &  & 2-0 & 3-1 & & 2-0 & 3-1 &  & \\ 
\midrule
HD 137613 & HdC & 0.174 & 0.127 & & 0.2 & 0.148 & $\sim$ 0.86 $\pm$ 0.02 & $>$ 15 \\ 
SV Sge & RCB & 0.46 & 0.45 & & 0.225 & 0.22 & $\geq$ 2.05 $\pm$ 0.01 & $>$ 45 \\ 
ES Aql & RCB & 0.373 & 0.362 & & 0.093 & 0.088 &  $\geq$ 4 $\pm$ 0.1 & $>$ 37 \\ 
DY\,Persei & DY\,Persei & 0.24 & 0.19 & & 0.06 & 0.045 & $\geq$ 4 $\pm$ 0.2 & $>$ 24 \\ 
ASAS J045331+2246.5 & DY\,Per suspect & 0.25 & 0.22 & & 0.052 & 0.045 & $\geq$ 5 $\pm$ 0.2 & $>$ 19 \\ 
V1368 Ori & DY\,Per suspect & 0.275 & 0.25 & & 0.05 & 0.045 & $\geq$ 5.5 $\pm$ 0.1 & $>$ 25 \\ 
EV Eri & DY\,Per suspect & 0.29 & 0.22 & & 0.04 & 0.03 & $\geq$ 7.5 $\pm$ 0.2 & $>$ 20 \\ 
CGCS 1049 & DY\,Per suspect & 0.25 & 0.23 & & 0.025 & 0.024 & $\geq$ 10 $\pm$ 0.5 & $>$ 19 \\ 

IRAS 05301+1805 & DY\,Per suspect & 0.24 & 0.23 & & \nodata & \nodata & \nodata & $>$ 19 \\
ASAS J065113+0222.1 & DY\,Per type star & 0.22 & 0.20 & & \nodata & \nodata & \nodata & $>$ 15 \\
\hline

\end{tabular}
\end{center}
\end{table*}

\begin{table*}[ht]
\caption{ Absorption depths of first overtone CO band heads and the estimated $^{12}$C/$^{13}$C ratios  of normal and cool carbon giants.} 
\begin{center}

\begin{tabular}{lllllll}\hline\label{Table.4}

Star name  & \multicolumn{2}{c}{$^{12}$C$^{16}$O} & & \multicolumn{2}{c}{$^{13}$C$^{16}$O}&  $^{12}$C/$^{13}$C  \\  
\cmidrule{2-3} \cmidrule{5-6} 
 &  2-0 & 3-1 & & 2-0 & 3-1 &  \\ 
\midrule
Arcturus & 0.228 & 0.205 & & 0.098 & 0.095 & $>$ 2.25  $\pm$ 0.2  \\ 

HD 156704 & 0.125 & 0.11 & & 0.04 & 0.032 & $>$ 3.25 $\pm$ 0.2 \\ 

HD 112127 & 0.215 & 0.225 & & 0.060 & 0.08 &  $>$ 3.2 $\pm$ 0.4 \\ 

BD+062063 & 0.255 & 0.268 & & 0.126 & 0.125 & $>$ 2.05 $\pm$ 0.1 \\ 

HR 337 & 0.24 & 0.21 & & 0.089 & 0.087 & $>$ 2.55 $\pm$ 0.2 \\ 

HD 64332 & 0.33 & 0.332 & & 0.158 & 0.165 & $>$ 2.05 $\pm$ 0.1 \\ 

HD 123821 & 0.167 & 0.186 & & 0.052 & 0.083 & $>$ 2.75 $\pm$ 0.5 \\ 

HR 3639 & 0.33 & 0.331 & & 0.16 & 0.145 & $>$ 2.15 $\pm$ 0.2 \\ 

HD 58521 & 0.365 & 0.322 & & 0.13 & 0.102 & $>$ 3 $\pm$ 0.2 \\ 

HD 76846  & 0.184 & 0.186 & & 0.088 & 0.101 & $>$ 1.9 $\pm$ 0.2 \\ 

V455\,Pup & 0.266 & 0.243 & & 0.066 & 0.07 & $>$ 3.75 $\pm$ 0.3 \\ 

TU\,Gem & 0.312 & 0.293 & & 0.068 & 0.075 & $>$ 4.2 $\pm$ 0.3 \\ 

Y\,CVn & 0.262 & 0.2512 & & 0.1632 & 0.16 & $>$ 1.6 $\pm$ 0.2 \\ 

RY\,Dra & 0.215 & 0.213 & & 0.123 & 0.118 & $>$ 1.75 $\pm$ 0.1 \\ \hline

\end{tabular}
\end{center}
\end{table*}
\section{PRELIMINARY RESULTS AND DISCUSSION}
The observed stars show strong 1st overtone bands of $^{12}$C$^{16}$O in the $K$-band region (see Figures \ref{fig.3} and \ref{fig.4}). 
As reported by \citet{hz-C2007}, prominent 1st overtone bands of $^{12}$C$^{18}$O are seen with no detection of $^{13}$C$^{16}$O in the two cool RCBs, SV\,Sge and ES\,Aql, and in the HdC star HD\,137613 (see Figure \ref{fig.3}); Z\,UMi spectrum is particularly noisy but suggests the presence of $^{12}$C$^{18}$O bands.
As expected, a close inspection of the $K$-band spectra of the observed normal cool giants clearly show the presence of $^{13}$C$^{16}$O bands including the prominent $^{12}$C$^{16}$O bands with no detection of $^{12}$C$^{18}$O bands (see Figure \ref{fig.4}). We have used these HdC/RCBs and cool giants spectra as comparisons to look for the detection of $^{12}$C$^{18}$O and $^{13}$C$^{16}$O bands in the observed spectra of DY\,Persei, DY\,Per type stars and DY\,Per suspects.

Among the DY\,Persei and seven DY\,Per affiliated stars we find suggestion of $^{12}$C$^{18}$O bands with no clear detection of $^{13}$C$^{16}$O bands in five of these stars : DY\,Persei, EV\,Eri, V1368\,Ori, ASAS J045331+2246.5, and CGCS\,1049 (see Figure \ref{fig.3}).
In Figure \ref{fig.3} spectra of two stars, ASAS J065113+0222.1 and IRAS\,05301+1805, do not show any suggestion of $^{12}$C$^{18}$O and $^{13}$C$^{16}$O bands within the detection limit. In the case of V1942\,Sgr's spectrum (see Figure \ref{fig.3}), numerous features are observed and we could not confirm the presence or absence of both $^{12}$C$^{18}$O and $^{13}$C$^{16}$O bands.

Based on the observed $K$-band spectra of HdC/RCBs, DY\,Persei, and DY\,Per affiliated stars, an attempt is made to estimate $^{16}$O/$^{18}$O values by measuring the absorption depths of $^{12}$C$^{16}$O and $^{12}$C$^{18}$O band heads using 2-0 as well as 3-1 bands. This exercise is more difficult for the DY\,Per type stars since spectra of cool stars are full of absorption features and the blending of these features with the identified $^{12}$C$^{18}$O band heads (in such low resolution spectra) is surely
a possibility. Yet with the exact wavelength matches we could confirm the presence of $^{12}$C$^{18}$O bands. As these bands are not completely resolved and the bands from the more abundant isotopic species are possibly saturated, the estimated $^{16}$O/$^{18}$O values are the lower limits in most cases (see Table \ref{Table.3}). Using synthetic spectra for the analysis is avoided as it is extremely difficult to identify all the contributing features from the observed low resolution spectra.

As all the DY\,Per affiliate stars observed here were reported to show strong presence of $^{13}$C in their respective discovery papers \citep{hz-M2012,hz-T2013}, we expected enhanced $^{13}$C$^{16}$O depths in the $K$-band spectra. We have estimated $^{12}$C/$^{13}$C ratios from the $K$-band absorption depths of $^{12}$C$^{16}$O and $^{13}$C$^{16}$O band heads. Since the observed depth at $^{13}$C$^{16}$O band heads is more or less comparable with the noise levels of the observed spectra, we conclude that there is no clear suggestion of $^{13}$C$^{16}$O in their spectra within the detection limit. However, we have estimated the lower limits of the $^{12}$C/$^{13}$C ratios measured from the $K$-band spectra of these stars as given in Table \ref{Table.3}. The depth at $^{13}$C$^{16}$O 2-0 band head region is used due to the better signal than other regions.  We find that our estimated lower limit on $^{12}$C/$^{13}$C for DY\,Persei is in line with the range of values, 20-50, obtained by \citet{hz-KB1997}.

 We have also estimated the $^{12}$C/$^{13}$C ratios for the observed normal and the cool carbon giants (see Table \ref{Table.4}) for comparison. These very low lower limits on $^{12}$C/$^{13}$C ratios measured for these carbon giants clearly show enhanced $^{13}$C in contrast to the DY\,Per affiliates. The $^{12}$C/$^{13}$C ratios are expected to be more than the estimated lower limits for the normal and cool carbon giants.

In the $H$-band region, the observed spectra do show the second overtone bands of $^{12}$C$^{16}$O but most of these are affected by noise. The strength of $^{12}$C$^{16}$O features in $H$-band is much weaker compared to those in $K$-band. Hence, detection of $^{12}$C$^{18}$O and $^{13}$C$^{16}$O in the $H$-band spectra is extremely difficult due to noise issues. For example, Figures \ref{fig.1} and \ref{fig.2} show
the atomic features as well as the wavelength positions of $^{12}$C$^{18}$O and $^{13}$C$^{16}$O band heads.

\section{CONCLUSIONS}

Our analysis  show the presence of strong $^{12}$C$^{18}$O band heads in RCB and HdC stars. The HdC star, HD\,137613, and the two RCB stars: SV\,Sge and ES\,Aql, are common with \citet{hz-C2007}.
Our $^{16}$O/$^{18}$O estimates for these three stars are in fair agreement with the values given in column (4) of \citet{hz-C2007}'s Table 2.

For DY\,Persei and the relatively cooler DY\,Per affiliated stars, our conclusion are less clear, however, there seems to be indication of $^{18}$O in the atmosphere of DY\,Persei and 4 DY\,Per suspects and no $^{13}$C (within the detection limit) which is the main isotopic signature of RCB/HdC stars. In the case of the DY\,Per type star, V1942\,Sgr, numerous features are observed and we could not confirm the presence or absence of both $^{12}$C$^{18}$O and $^{13}$C$^{16}$O bands. Note that, the $K$-band spectra of all the normal carbon stars, with similar S/N spectra of DY\,Per affiliates, having similar effective temperatures, show prominent $^{13}$C$^{16}$O bands. 
On the contrary, one DY\,Per type star ASAS J065113+0222.1, and one DY\,Per suspect IRAS\,05301+1805 show little or no presence of both $^{18}$O and $^{13}$C in their atmosphere.
  
So whether DY\,Per type stars are the cooler cousins of RCBs or just a counterpart of normal carbon rich AGBs suffering ejection events can be better explored through the analyses of high resolution $H$- and $K$-band spectra. Our preliminary analysis suggests that a quartet of suspects along with DY\,Persei itself show prominent $^{12}$C$^{18}$O bands and no $^{13}$C$^{16}$O bands, which is in sharp contrast to the normal  carbon stars and much similar to RCBs, and builds up a strong case to dig deeper into the high resolution spectra of these stars to find their evolutionary origins.
\acknowledgments
It is our pleasure to thank the referee for a constructive report that helped us considerably in the presentation of this work. We would like to thank the staff at IAO, Hanle and the remote control station  at  CREST,  Hosakote for assisting in observations. We thank Dr. J. P. Ninan for his valuable suggestions regarding observations and reductions. We also thank Prof. Rajat Chowdhury for giving us valuable inputs in calculating the isotopic shifts.\\

\bibliographystyle{apj}



\end{document}